\documentclass[runningheads]{llncs}
\pdfoutput=1 
\usepackage[T1]{fontenc}

\usepackage{graphicx}
\usepackage{amsmath}
\usepackage{hyperref}
\usepackage{subfigure}
\usepackage{multirow}
\usepackage{booktabs}
\usepackage{array}
\usepackage{float}
\usepackage{booktabs}
\usepackage{pifont}
\usepackage{orcidlink}
\usepackage{marvosym}

\begin{document} 
\title{Fundus2Video: Cross-Modal Angiography Video Generation from Static Fundus Photography with Clinical Knowledge Guidance}
\titlerunning{Fundus2Video}
\author{Weiyi Zhang\inst{1}\orcidlink{0009-0008-2780-9121} \and
Siyu Huang\inst{2} \and
Jiancheng Yang\inst{3} \and
Ruoyu Chen \inst{1} \and
Zongyuan Ge \inst{4} \and
Yingfeng Zheng \inst{5} \and
Danli Shi \inst{1}  \textsuperscript{(\Letter)} \and
Mingguang He \inst{1}}
%\author{Anonymous}
%
\authorrunning{Zhang et al.}
%\authorrunning{Anonymous}
%\institute{Anonymous Organization}
\institute{The Hong Kong Polytechnic University, Kowloon, Hong Kong \and
Clemson University, South Carolina, USA \and
École Polytechnique Fédérale de Lausanne (EPFL), Lausanne, Switzerland \and
Monash University, Melbourne, Australia \and
Sun Yat-sen University, Guangzhou, China\\
\email{danli.shi@polyu.edu.hk}}
\maketitle              % typeset the header of the contribution
\begin{abstract}
Fundus Fluorescein Angiography (FFA) is a critical tool for assessing retinal vascular dynamics and aiding in the diagnosis of eye diseases. However, its invasive nature and less accessibility compared to Color Fundus (CF) images pose significant challenges. Current CF to FFA translation methods are limited to static generation. In this work, we pioneer dynamic FFA video generation from static CF images. 
We introduce an autoregressive GAN for smooth, memory-saving frame-by-frame FFA synthesis. To enhance the focus on dynamic lesion changes in FFA regions, we design a knowledge mask based on clinical experience. Leveraging this mask, our approach integrates innovative knowledge mask-guided techniques, including knowledge-boosted attention, knowledge-aware discriminators, and mask-enhanced patchNCE loss, aimed at refining generation in critical areas and addressing the pixel misalignment challenge. Our method achieves the best FVD of 1503.21 and PSNR of 11.81 compared to other common video generation approaches. Human assessment by an ophthalmologist confirms its high generation quality. Notably, our knowledge mask surpasses supervised lesion segmentation masks, offering a promising non-invasive alternative to traditional FFA for research and clinical applications. 
The code is available at \url{https://github.com/Michi-3000/Fundus2Video}.
\end{abstract}
\section{Introduction}
Fundus Fluorescein Angiography (FFA) is an essential examination in ophthalmology clinics, providing a dynamic view of retinal blood flow and lesion changes. It offers critical insights into retinal circulatory dynamics, aiding in the identification of conditions such as diabetic retinopathy, hypertensive retinopathy, and macular degeneration \cite{impffa}. Unlike Color Fundus (CF) images, FFA videos capture the dynamic filling process and real-time changes in retinal vascular abnormalities with greater clarity and depth, thereby enhancing diagnostic precision and facilitating a deeper understanding of disease progression and treatment response. However, due to its invasive nature and potential side effects, FFA's use is limited for certain individuals. In contrast, CF photography is non-invasive, readily available \cite{fundus1}, and has been utilized in some deep-learning methods \cite{detec1,detec2} for disease diagnosis. Therefore, generating realistic FFA videos from CF images holds significant research and application potential.

When considering the generative models for FFA synthesis, the majority of existing methods \cite{FFA1,FFA3,vs1,vs2,ours} focus on specific phases, like the venous and late phase, using various Generative Adversarial Networks (GAN). However, they overlook the changes occurring throughout the entire FFA process, which includes multiple phases. While some approaches \cite{3slice} can generate multiple discrete FFA images from different phases simultaneously, they still cannot capture the fully dynamic changes of retinal structures and lesions. Capturing lesional changes accurately is another challenge in FFA generation. 
While using lesion labels for conditional supervision could potentially enhance image details, the manual annotation of these labels is highly time-consuming and impractical for segmenting all possible lesion changes.
Additionally, the time-consuming nature of FFA procedures makes it difficult to align FFA images precisely with CF images in clinical practice, due to blinking and movement, even with good patient cooperation \cite{fundus2,fundus4}. This misalignment poses a significant challenge for pixel-to-pixel-based video generation processes. 

To tackle these challenges, we propose a model leveraging an image-to-image GAN framework, specifically pix2pixHD \cite{pix2pixHD}, to generate smooth and stable FFA videos from single CF images autoregressively. 
Through clinical knowledge analysis of ground-truth FFA series, regions with significant lesion changes during the early and late FFA series examination are lesional changes, reflecting the damage in vascular or retinal pigment epithelium structure \cite{theory,fundus3}. The larger the changes, the more important they are. Leveraging this insight, we design a knowledge mask that requires no additional manual labeling and enhances the generation of regions with high variability. Using this mask, we introduce novel knowledge mask-guided techniques into the baseline model to guide the model to focus more on key regions during learning and generation. Specifically, we propose a mask-enhanced patchNCE loss to address the pixel misalignment issue. This model holds the potential to generate FFA videos from CF images to other modalities and improve downstream tasks \cite{icga,song2023deep,SHI2024100363}.

In summary, our research contributes as follows:
\textbf{1.} We are the first to generate dynamic FFA video directly from CF images, marking a significant advancement in ophthalmic imaging. Specifically, we introduce Fundus2Video, an autoregressive GAN architecture tailored for frame-by-frame FFA video synthesis from CF images. This architecture optimizes memory usage and ensures smooth output. \textbf{2.} We introduce a knowledge mask derived from clinical insights to enhance focus on regions undergoing significant changes during dynamic FFA processes. This eliminates manual labeling and improves generation in areas like lesions and blood vessels. \textbf{3.} With this mask, we implement knowledge mask (KM)-guided techniques. We introduce knowledge-boosted attention and knowledge-aware discriminators for specific supervision on regions of lesion regions. To address the pixel misalignment challenge between CF images and ground-truth FFA series in critical areas, we employ a newly designed mask-enhanced patchNCE loss.

\begin{figure*}[tb]
	\centering
	\includegraphics[width=12cm, height=5.8cm]{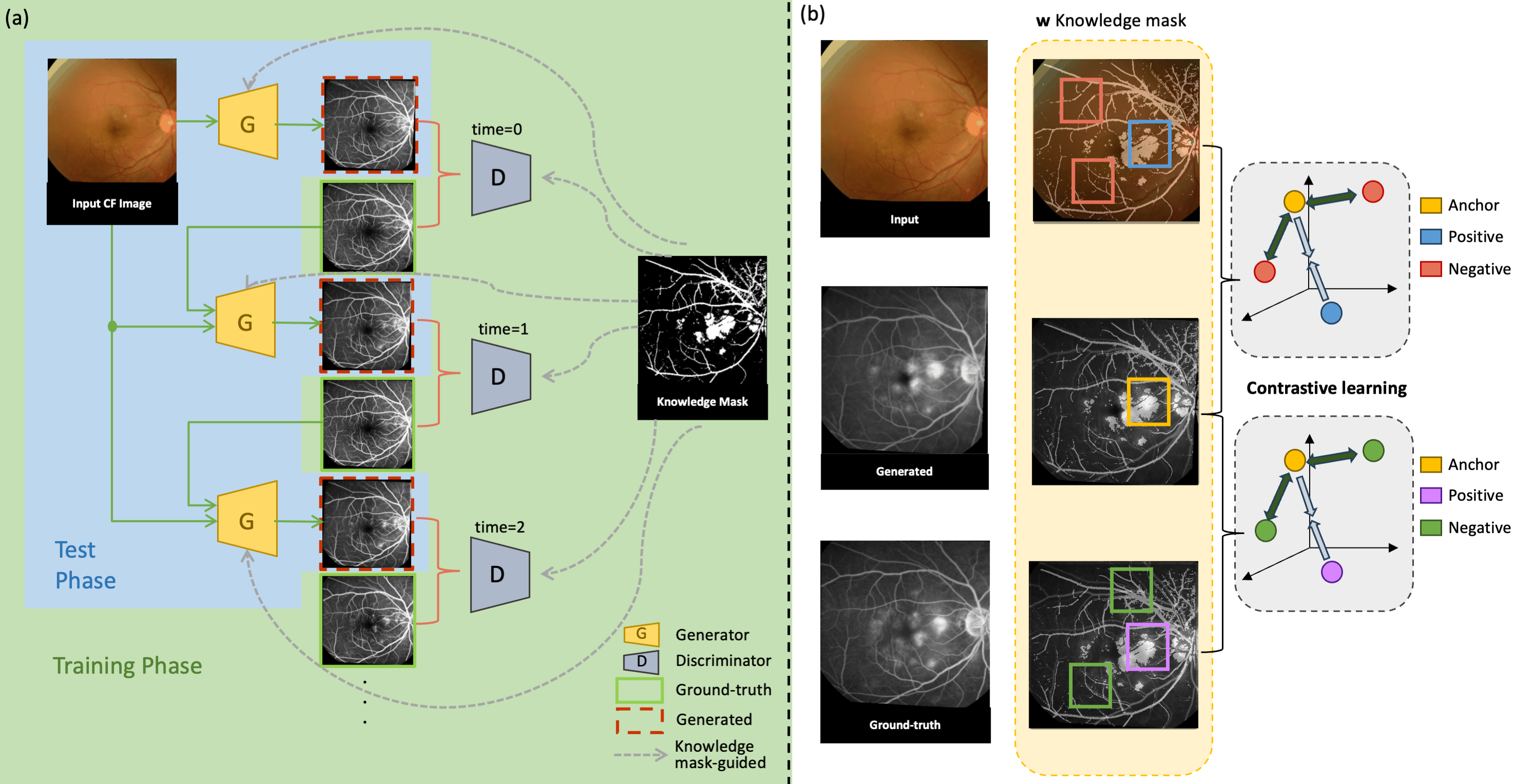}
	\caption{Proposed Fundus2Video. (a) The overall architecture. Generator $G$ generates one frame at a time, taking the output from the previous time step and the CF image as input. During the training phase, unsupervised knowledge masks guide the entire network. (b) The design of the mask-enhanced patchNCE loss. }
	\label{fig:2}
\end{figure*}

\section{Methods}
%Our Fundus2FA 
\subsection{Overview}
We aim to generate a realistic FFA video \( \hat{Y} \) from a given CF image \( x \), with the ground-truth FFA video during training represented as \( Y \).
Considering the temporal nature of FFA series, we adopt an autoregressive GAN architecture to capture temporal dependencies and generate coherent video sequences. An autoregressive GAN generates image samples sequentially, conditioning each new image on previously generated images and additional inputs. In our context of generating FFA videos from CF images, our autoregressive GAN, named Fundus2Video, based on the image-to-image translation GAN pix2pixHD, sequentially generates each frame of the FFA video, incorporating the CF image itself and the preceding frames. Building upon the generator, discriminator, and loss designs of pix2pixHD, our approach incorporates specific modifications to enable autoregressive and smooth generation. The architecture is as shown in Fig. \ref{fig:2} (a).

To ensure smooth output in Fundus2Video, we incorporate multi-frame input and smoothing techniques for longer temporal considerations. 
Specifically, we input three consecutive frames from the ground-truth FFA series to the model in a sliding window fashion to provide longer temporal context for each generated frame. Instead of generating each frame independently, we aggregate the generated frames over a sliding window and perform triple-frame averaging. This approach smooths out abrupt transitions between frames and ensures continuity in the generated video sequence.

\begin{figure}[t]
	\centering
	\includegraphics[width=11cm, height=3cm]{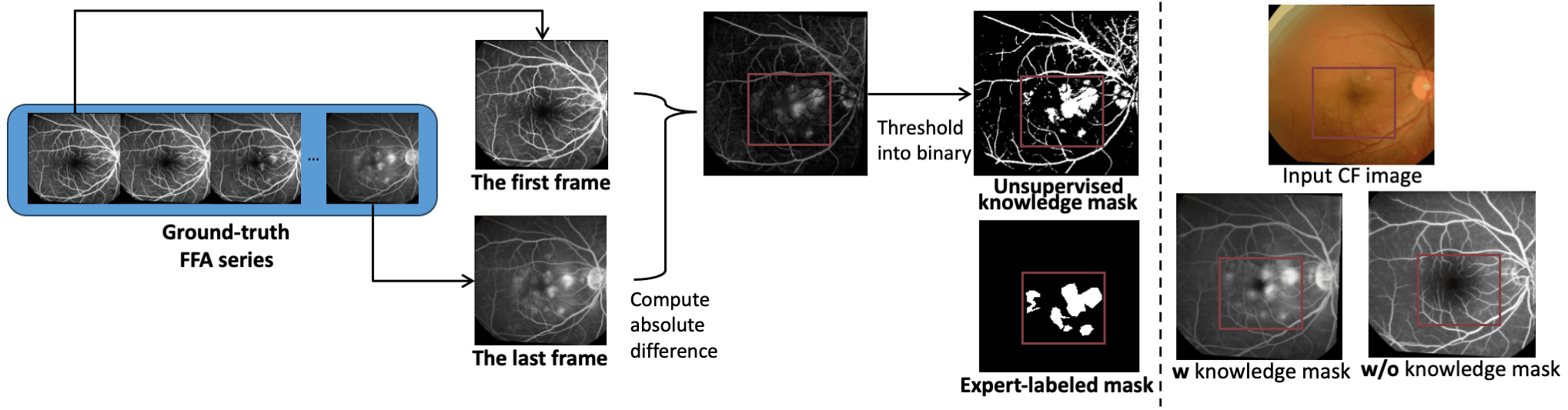}
	\caption{The definition of the knowledge mask. Left: The unsupervised process of obtaining the mask. The knowledge mask covers the same pathological areas as the expert-labeled mask. Right: Generated results with and without the knowledge mask.}
	\label{fig:1}
\end{figure}
\subsection{Unsupervised Clinically Supported Knowledge Mask}
The baseline Fundus2Video can generate smooth and continuous FFA videos. However, it falls short of accurately depicting details like lesions and critical structures as shown in Fig. \ref{fig:1} right, which are of utmost clinical importance. To address this, we leveraged clinical insights to analyze ground-truth FFA videos, which tell us regions undergoing significant morphological changes during the FFA process often corresponded to crucial lesions or retinal structure areas that pose challenges for the model. The theoretical basis is from \cite{theory}: 
\begin{itemize}
    \item \textit{During the FFA process, as the fluorescent dye flows through retinal vessels, significant leakage always occurs around the lesions, leading to visible differences between early and late stages.} 
\end{itemize}
Building upon this knowledge, we devised a simple binary mask by computing the difference between the first frame (representing the arterial phase) and the last frame (representing the late phase) and setting a specific threshold $\delta$ determined through comparative experiments, which can be formulated as \(m = \delta(Y_{\text{0}} - Y_{\text{T}})\), where \( Y_{\text{0}} \) represents the first frame of the ground-truth FFA video and \( Y_{\text{T}} \) represents the last frame. The process is depicted in Fig. \ref{fig:1}. Unlike supervised lesion/structure segmentation masks, this knowledge mask requires no additional manual annotation or segmentation model training. It can be easily derived from raw data, making it simple yet effective.

\subsection{Knowledge Mask-Guided Video Generation}
%In this section, we introduce three knowledge mask (KM)-guided tech including three techniques to leverage knowledge masks for enhanced supervision.
\subsubsection{Knowledge-boosted Attention. }
Some types of lesions may be challenging to detect in CF images due to low contrast, leading to synthesized FFA slices lacking details in these areas. To address this limitation and improve the generator's ability to capture specific regions, we introduce additional supervision into the learning process. Our approach, termed knowledge-boosted attention, involves guiding the network's attention toward focal regions during training. To quantify this guidance, we define an attention loss $\mathcal{L}_{Att}$ as follows:

\begin{equation}
\mathcal{L}_\text{Att}(A,m)=\frac{1}{n}\sum_{i}^{}(A^{i}-m^{i})^2.
\end{equation}
Here, $m$ represents the knowledge mask described in Section 2.2. $A$ denotes the attention map obtained by element-wise multiplication of the semantic-rich activation map $f_l$ from the last convolutional layer $l$ in the generator and the mask $m$. We then apply a rectified linear operation to $A$, resulting in $A = ReLU(f_l\odot m)$.

\subsubsection{Mask-enhanced PatchNCE Losses.}
%In Section 3.1, we introduced patchNCE losses to mitigate data misalignment issues. 
To address pixel misalignment in ground-truth FFA series and CF images caused by motion artifacts during acquisition, we introduce the PatchNCE loss \cite{patchNCE2}, inspired by contrastive learning techniques known for boosting model robustness against label noise. 
%The PatchNCE loss maximizes mutual information between input, output, and ground truth for content consistency. 
However, we observed that the model's primary focus should be on reducing jitter in clinically relevant regions, such as lesions and vasculature, which are of greater clinical significance. To further tackle this issue, we propose mask-enhanced PatchNCE losses as a replacement for traditional PatchNCE losses. This method extends traditional PatchNCE losses by incorporating a knowledge mask \( m \), highlighting critical regions within the FFA series. 
%Mathematically, it minimizes an InfoNCE loss \cite{infoNCE} calculated patch-wise:
Mathematically, the proposed mask-enhanced PatchNCE losses are based on the InfoNCE loss, which is defined as:
\begin{equation}
  \mathcal{L}_\text{InfoNCE}(v, v^+, v^-) = -\log\left(\frac{e^{sim(v, v^+)}}{e^{sim(v, v^+)} + \sum_{j=1}^Ne^{sim(v, v^-_j)}}\right).
  \label{eq:patchNCE}
\end{equation}
Here, \( v \), \( v^+ \), and \( v^- \) represent the embeddings of the anchor, positive, and negative samples, respectively.

The mask-enhanced unsupervised PatchNCE (UP) loss compares the anchor patch \( z_{\hat{Y}} \) in the generated output with a corresponding positive patch \( z_X \) from the input CF image and negative patches \( z_X^- \), under the guidance of knowledge mask $m$. It is defined as: 
\begin{equation}
    L_\text{Masked\ UP} = L_\text{InfoNCE}(m \odot z_{\hat{Y}}, m \odot z_X, m \odot z_X^-) ,
\end{equation}
where \( \odot \) denotes element-wise multiplication. In contrast, the mask-enhanced supervised PatchNCE (SP) loss ensures consistency between generated and ground-truth patches. It designates the corresponding patch in the ground-truth image \( z_Y \) as positive, while non-corresponding patches \( z_Y^- \) are considered negatives. It's defined as:
\begin{equation}
    L_\text{Masked\ SP} = L_\text{InfoNCE}(m \odot z_{\hat{Y}}, m \odot z_Y, m \odot z_Y^-) .
\end{equation}

The illumination is shown in Fig. \ref{fig:2} (b). By integrating the knowledge mask into the PatchNCE loss, our method directs the model's focus during training, improving its ability to capture clinically significant features.

\subsubsection{Knowledge-aware Discriminators. }
We employ 3 discriminators $D=\{D_1,D_2,D_3\}$ \cite{iizuka2017globally,pix2pixHD} with the same patchGAN architecture \cite{pix2pix} to evaluate images at scales of 1, 0.5 and 0.25 for different receptive fields.
The discriminator objective function for $D_k$ with generator $G$ is given by:
\begin{equation}
\mathcal{L}_{D_k}(a,b,G(a))=E_{a,b}[logD_k(a,b)]+E_{a}[log(1-D_k(a,G(a)))]\label{con:loss1},
\end{equation}
where $a$, $b$ and $G(a)$ are the input, ground-truth, and generated images.

However, solely discriminating the entire image may not ensure the authenticity of lesion regions in generated FFA frames. Hence, we introduce discrimination guided by knowledge across scales. By combining knowledge masks $m$ with corresponding FFA images, we tailor inputs for the discriminators to focus on lesions. According to Eq. \ref{con:loss1}, the combined discriminator loss $L_{GAN}(G, D_{k})$ for scale $k$ is defined as:
\begin{equation}
\mathcal{L}_\text{GAN}(G,D_{k})=\mathcal{L}_{D_k}(x,y,G(x))+\mathcal{L}_{D_k}(x \odot m,y \odot m,G(x) \odot m),
\end{equation}
where $\odot$ denotes element-wise multiplication, and $x$ and $y$ are the input CF images and ground-truth FFA images, respectively.

Consequently, the final loss function is as follows: 
\begin{equation}
\mathcal{L} = \lambda_\text{UP}\mathcal{L}_\text{MaskedUP}+\lambda_\text{SP}\mathcal{L}_\text{MaskedSP}+\lambda_\text{Att}\mathcal{L}_\text{Att}+\lambda_\text{GAN}\mathcal{L}_\text{GAN}.
\label{con:lossf}
\end{equation}
%where $\lambda_1$, $\lambda_2$ and $\lambda_3$ are hyperparameters that adjust the respective loss weights. 

\section{Experiments}
\subsubsection{Dataset.}
Our dataset comprises 350 CF images and 18,180 corresponding FFA images from 350 anonymous patients sampled from a large paired dataset. The FFA images were obtained using Zeiss FF450 Plus and Heidelberg Spectralis systems, with a resolution of 768×768 pixels. Meanwhile, the CF images were captured by Topcon TRC-50XF and Zeiss FF450 Plus instruments, with resolutions ranging from 1,110×1,467 to 2,600×3,200 pixels. The Institutional Review Board approved the study.

\subsubsection{Implementation Details.}
The final objective function (Eq.~\ref{con:lossf}) was utilized to train the generative model, with $\lambda_{UP}$, $\lambda_{SP}$, $\lambda_{Att}$, and $\lambda_{GAN}$ set to 1, 1, 4, and 2, respectively. The threshold $\delta$ for obtaining the knowledge mask was set to 45.
During training, each ground-truth FFA series produced 12 frames, with 4 slices randomly selected from the vascular, venous, and late phases, respectively. Data augmentation techniques including random cropping, scaling, and color augmentation. The input images were resized to 512×512. Additionally, the model was trained to randomly select either generated or ground-truth frames as input, enhancing its adaptability and robustness. We employed the Adam optimizer with $beta_1=0.5$ and $beta_2=0.999$, adjusting the learning rate every 50 iterations using the PyTorch \cite{pytorch} lr-scheduler. The initial learning rate was set to 2e-3, with a batch size of 1. Training was conducted for 50 epochs on an NVIDIA GeForce RTX3090. For evaluation, 70$\%$ of the data was reserved for training at the patient level, while the remaining data was evenly split into validation and test sets.

\subsubsection{Evaluation Criteria.}
%FVD SSIM PSNR LPIPS
Our video evaluation criteria include Fréchet Video Distance (FVD) \cite{FVD}, Structural Similarity Index (SSIM) \cite{ssim}, Peak Signal-to-Noise Ratio (PSNR) \cite{psnr}, and Learned Perceptual Image Patch Similarity (LPIPS) \cite{LPIPS}. 
They measure feature distribution similarity, video structural similarity, reconstruction quality, and perceptual similarity, respectively.

	\begin{figure}[!t]%[htbp]
	\centering
	\begin{tabular}{m{1cm}<{\centering}m{1.5cm}<{\centering} m{1.5cm}<{\centering}m{1.5cm}<{\centering} m{1.5cm}<{\centering}m{1.5cm}<{\centering}m{1.5cm}<{\centering}m{1.5cm}<{\centering}}
		&\scriptsize{Input} & \scriptsize{Knowledge Mask}&\scriptsize{Frame 1}&\scriptsize{Frame 4} & \scriptsize{Frame 7}& \scriptsize{Frame 10} & \scriptsize{Frame 12}\\
		\centering
        GT&\includegraphics[width=1.5cm, height=1.5cm]{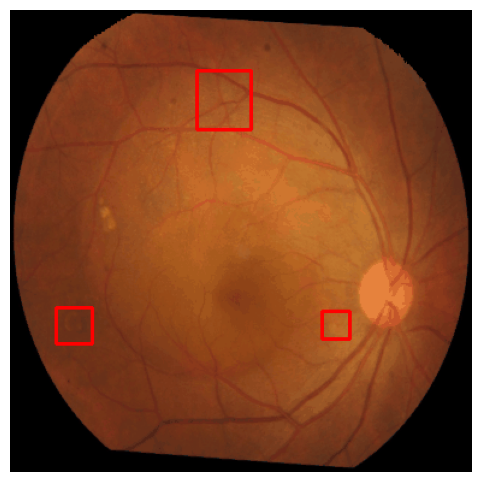} &
             \includegraphics[width=1.5cm, height=1.5cm]{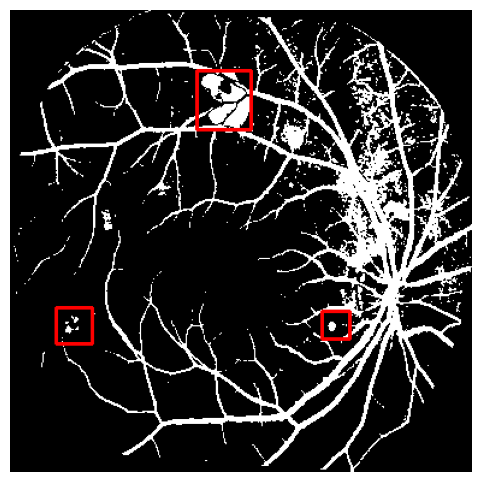} &
		\includegraphics[width=1.5cm, height=1.5cm]{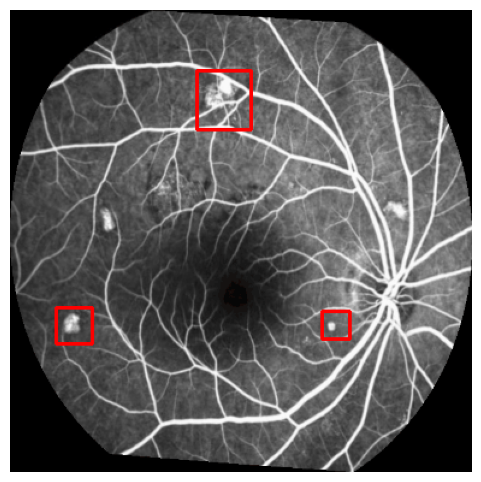} &
		\includegraphics[width=1.5cm, height=1.5cm]{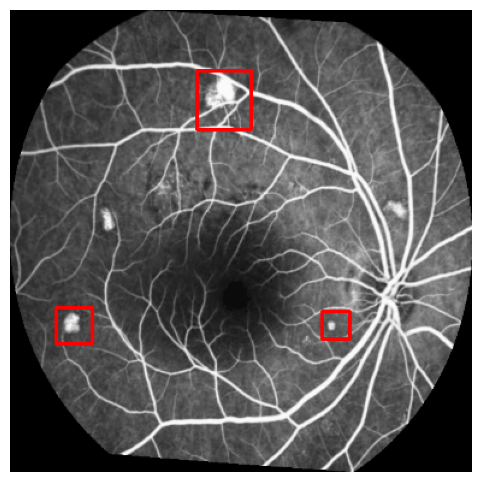} &
		\includegraphics[width=1.5cm, height=1.5cm]{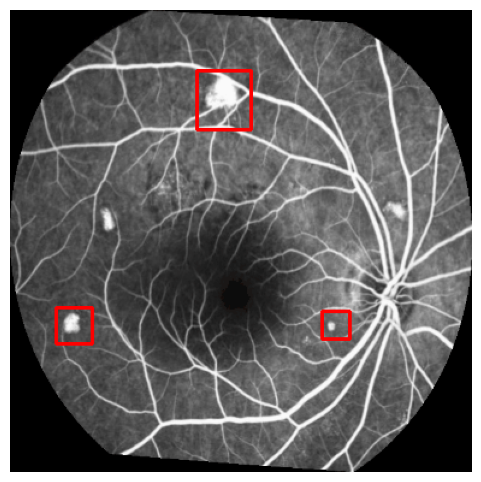} &
		\includegraphics[width=1.5cm, height=1.5cm]{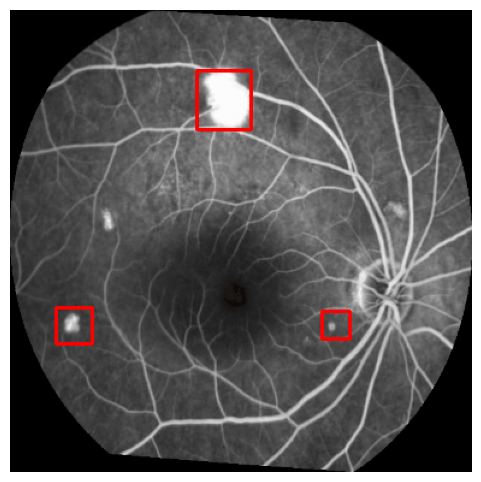} &
        \includegraphics[width=1.5cm, height=1.5cm]{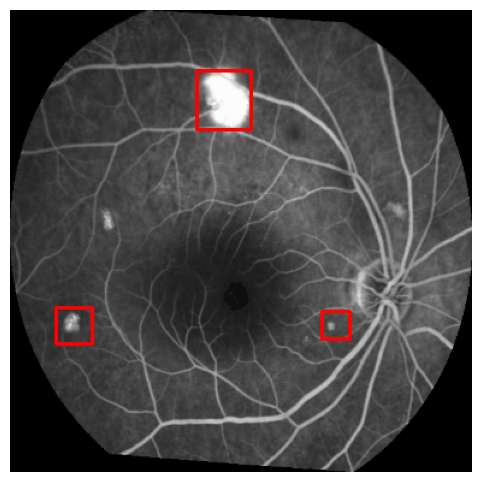} \\

        Seg2Vid&\includegraphics[width=1.5cm, height=1.5cm]{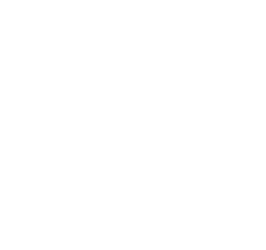} &
             \includegraphics[width=1.5cm, height=1.5cm]{blank.png} &		\includegraphics[width=1.5cm, height=1.5cm]{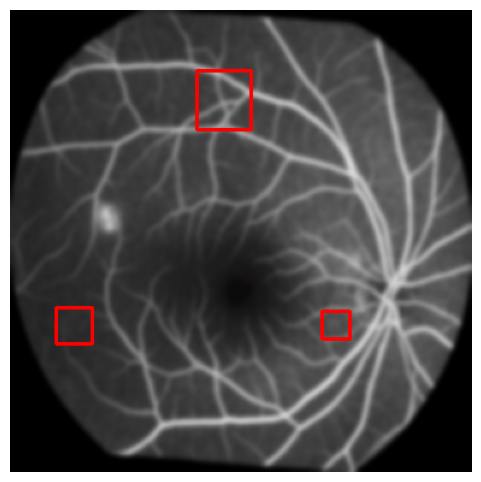} &
		\includegraphics[width=1.5cm, height=1.5cm]{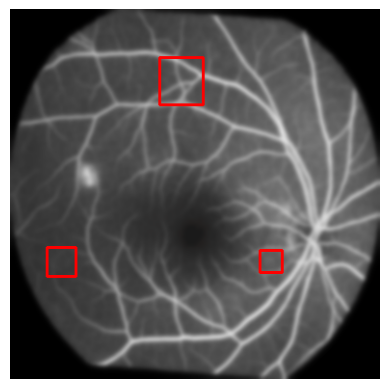} &
        \includegraphics[width=1.5cm, height=1.5cm]{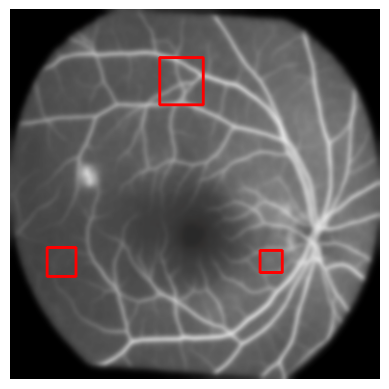} &
		\includegraphics[width=1.5cm, height=1.5cm]{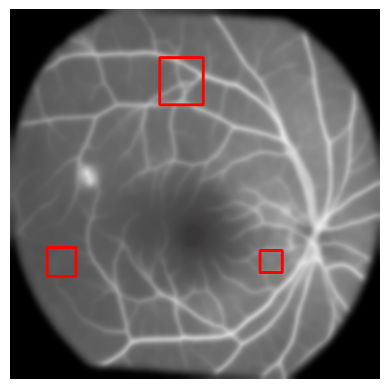} &
		\includegraphics[width=1.5cm, height=1.5cm]{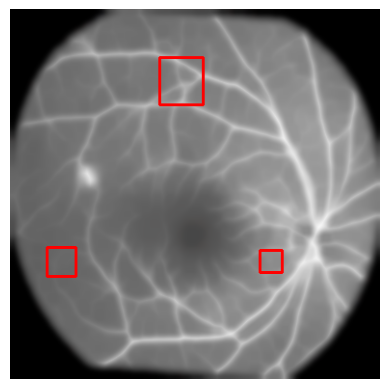} \\

         Med-ddqm&\includegraphics[width=1.5cm, height=1.5cm]{blank.png} &
             \includegraphics[width=1.5cm, height=1.5cm]{blank.png} &		\includegraphics[width=1.5cm, height=1.5cm]{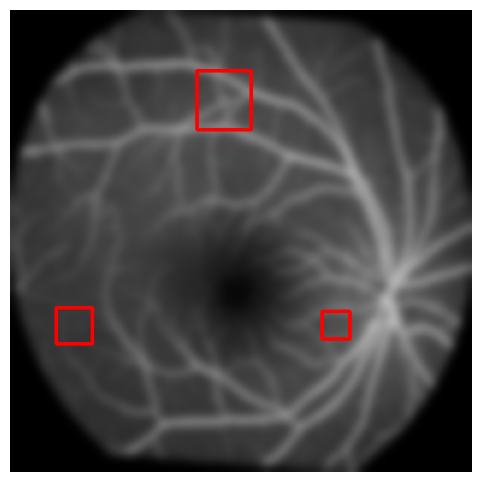} &
		\includegraphics[width=1.5cm, height=1.5cm]{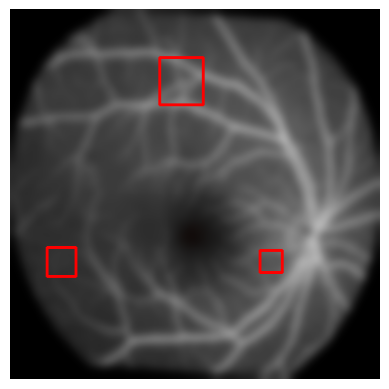} &
        \includegraphics[width=1.5cm, height=1.5cm]{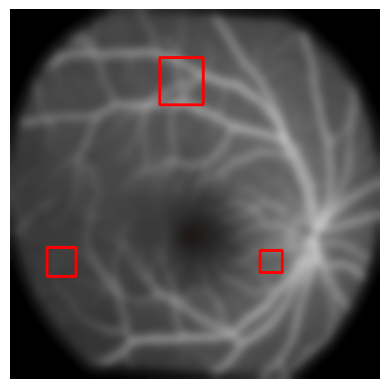} &
		\includegraphics[width=1.5cm, height=1.5cm]{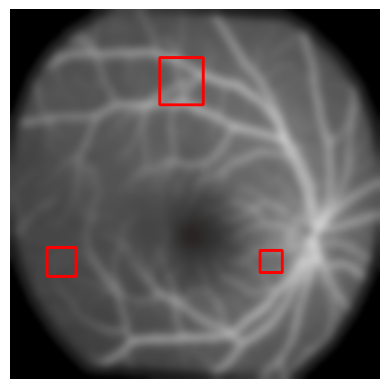} &
		\includegraphics[width=1.5cm, height=1.5cm]{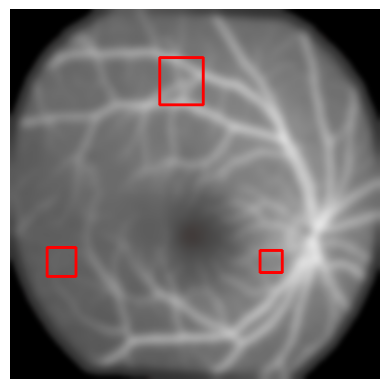} \\

        ConsistI2V&\includegraphics[width=1.5cm, height=1.5cm]{blank.png} &
             \includegraphics[width=1.5cm, height=1.5cm]{blank.png} &		\includegraphics[width=1.5cm, height=1.5cm]{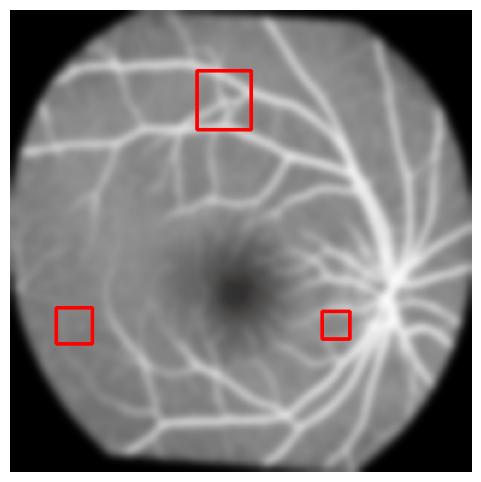} &
		\includegraphics[width=1.5cm, height=1.5cm]{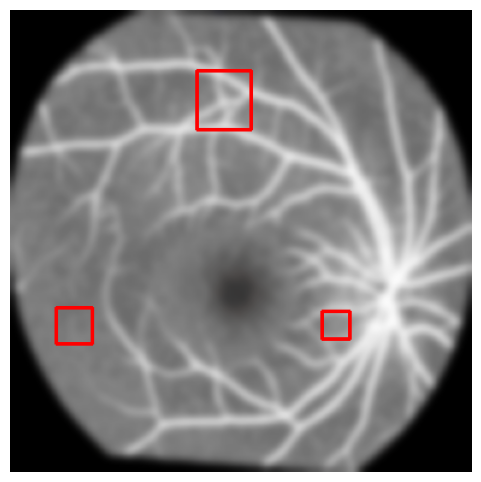} &
        \includegraphics[width=1.5cm, height=1.5cm]{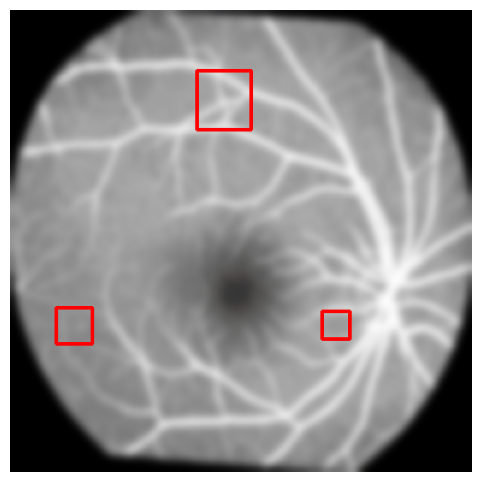} &
		\includegraphics[width=1.5cm, height=1.5cm]{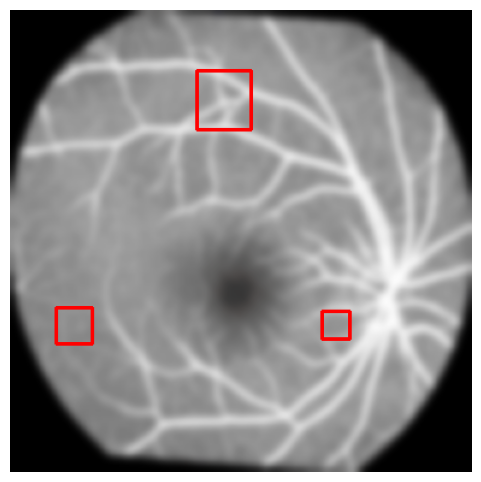} &
		\includegraphics[width=1.5cm, height=1.5cm]{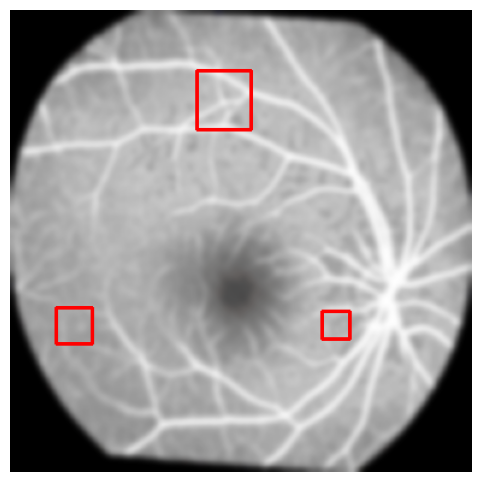} \\
		
		Fundus2Video\newline {w/o KM}&\includegraphics[width=1.5cm, height=1.5cm]{blank.png} &
             \includegraphics[width=1.5cm, height=1.5cm]{blank.png} &		\includegraphics[width=1.5cm, height=1.5cm]{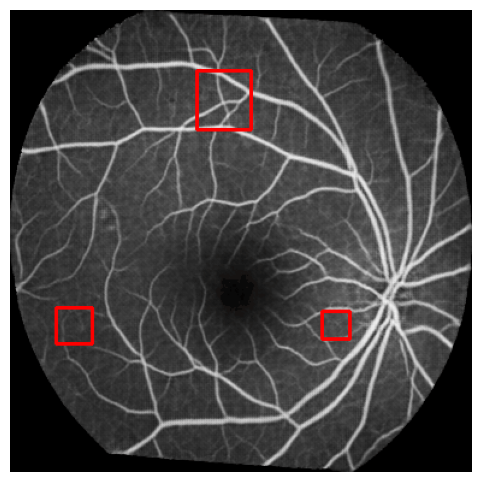} &
		\includegraphics[width=1.5cm, height=1.5cm]{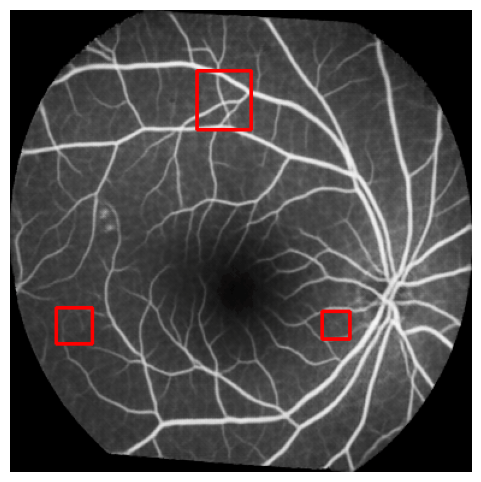} &
        \includegraphics[width=1.5cm, height=1.5cm]{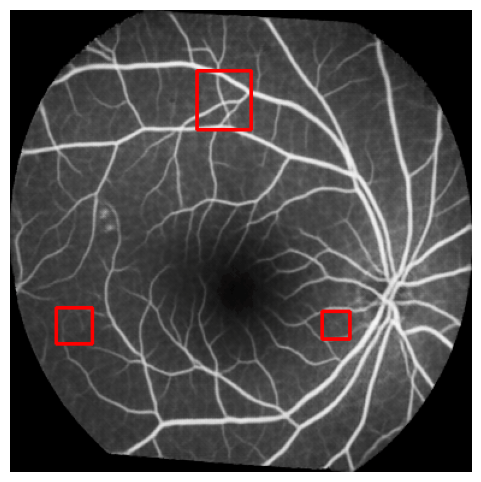} &
		\includegraphics[width=1.5cm, height=1.5cm]{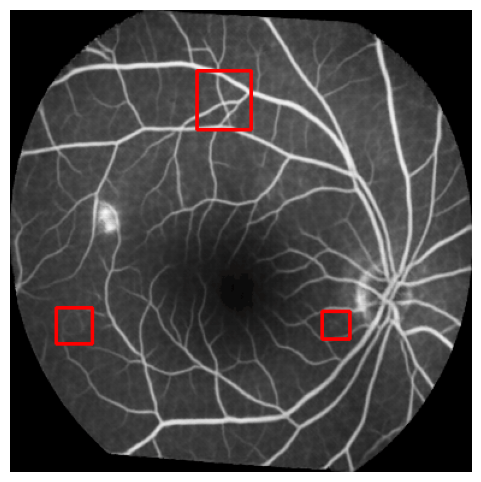} &
		\includegraphics[width=1.5cm, height=1.5cm]{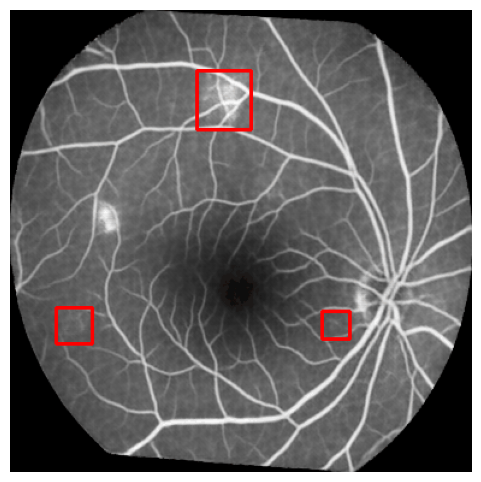} \\
		
		Fundus2Video&\includegraphics[width=1.5cm, height=1.5cm]{blank.png} &
             \includegraphics[width=1.5cm, height=1.5cm]{blank.png} &		\includegraphics[width=1.5cm, height=1.5cm]{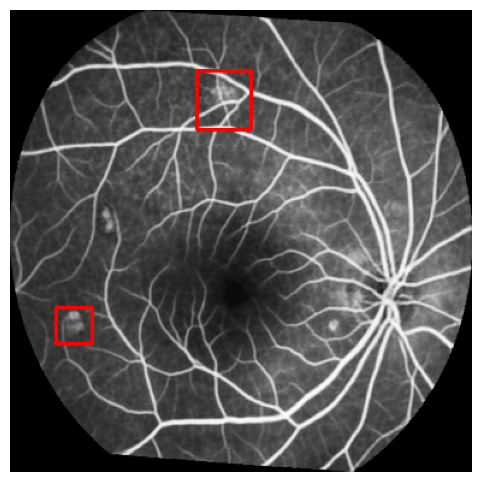} &
		\includegraphics[width=1.5cm, height=1.5cm]{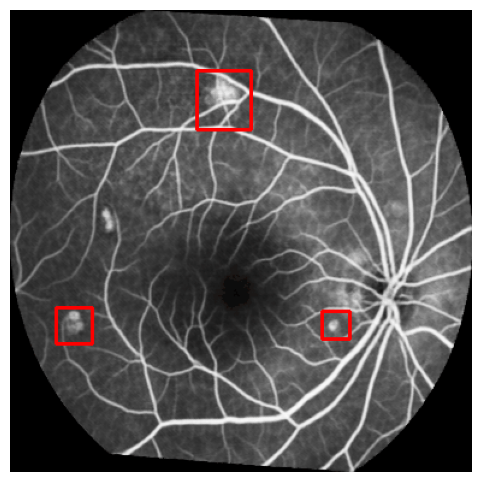} &
		\includegraphics[width=1.5cm, height=1.5cm]{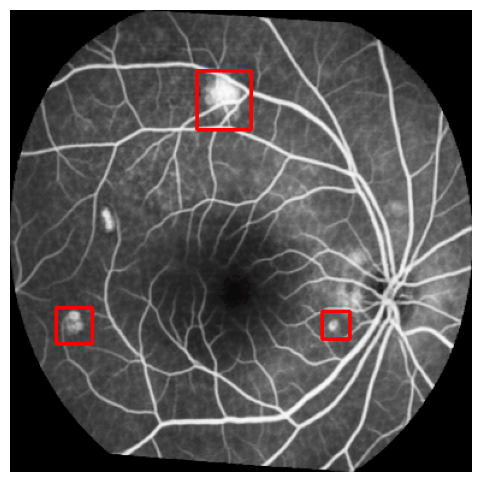} &
		\includegraphics[width=1.5cm, height=1.5cm]{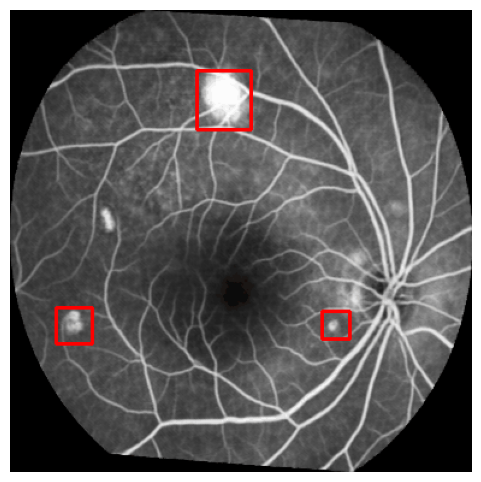}&
		\includegraphics[width=1.5cm, height=1.5cm]{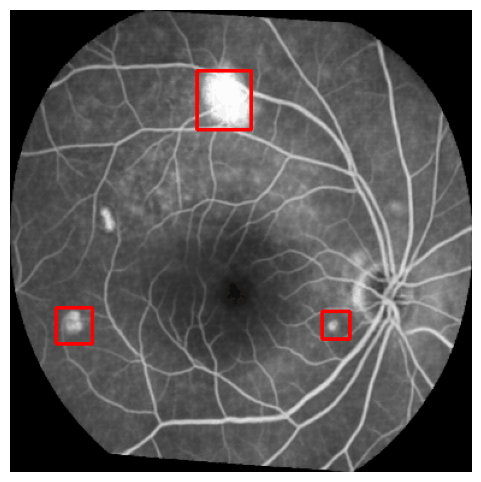} \\
		\end{tabular}
		\caption{Qualitative comparison of the methods. Frames are sampled from the 12-frame video. Areas in red boxes denote significant lesions. It can be observed that the KM-guided Fundus2Video exhibits best performance in generating critical lesions.}
		\label{fig:result}
	\end{figure}
\subsubsection{Model Comparisons.}
We evaluate Fundus2Video against existing image-to-video translation methods, including the auto-encoder-based Seg2vid \cite{seg2vid}, and the diffusion model-based Med-ddpm \cite{med_dif} and ConsistI2V \cite{ren2024consisti2v}. Table~\ref{tab:t1} shows our model's superior performance across all metrics. Qualitative comparison in Fig.~\ref{fig:result} reveals clearer images and discernible lesion areas in our approach versus others.
\begin{center}
\begin{table*}[t]
\centering
\caption{Comparison of the methods. $M$ stands for masks.}
\resizebox{1\textwidth}{!}{
\begin{tabular}{clccclccccc}
\hline
\multicolumn{1}{l}{\multirow{2}{*}{\textbf{Models}}} &
  \multirow{2}{*}{\textbf{Mask Type}} &
  \multicolumn{5}{c}{\textbf{Proposed Techniques}} &
  \multirow{2}{*}{\textbf{FVD$\downarrow$}} &
  \multirow{2}{*}{\textbf{SSIM$\uparrow$}} &
  \multirow{2}{*}{\textbf{PSNR$\uparrow$}} &
  \multirow{2}{*}{\textbf{LPIPS$\downarrow$}} \\
\multicolumn{1}{l}{} &
   &
  \multicolumn{1}{l}{\begin{tabular}[c]{@{}l@{}}PatchNCE\\ loss $L_{P}$\end{tabular}} &
  \multicolumn{1}{l}{\begin{tabular}[c]{@{}l@{}}Mask-enhanced\\ PatchNCE loss \\ $L_{\text{Masked}\ P}$\end{tabular}} &
  \multicolumn{2}{l}{\begin{tabular}[c]{@{}l@{}}Knowledge\\ -boost\\ attention $\mathcal{L}_{\text{Att}}$\end{tabular}} &
  \multicolumn{1}{l}{\begin{tabular}[c]{@{}l@{}}Knowledge\\ -aware\\ discriminators\end{tabular}} &
   &
   &
   &
   \\ \hline
\multicolumn{1}{l}{\textbf{Seg2Vid\cite{seg2vid}}} &
  \multicolumn{1}{c}{-} &
  - &
  - &
  \multicolumn{2}{c}{-} &
  - &
  2302.15 &
  0.2930 &
  10.23 &
  0.2451 \\
\multicolumn{1}{l}{\textbf{Med-ddpm\cite{med_dif}}} &
  \multicolumn{1}{c}{-} &
  - &
  - &
  \multicolumn{2}{c}{-} &
  - &
  2410.54 &
  0.2305 &
  10.59 &
  0.2513 \\
\multicolumn{1}{l}{\textbf{ConsistI2V\cite{ren2024consisti2v}}} &
  \multicolumn{1}{c}{-} &
  - &
  - &
  \multicolumn{2}{c}{-} &
  - &
  2108.33 &
  0.2662 &
  10.71 &
  0.2498 \\ \hline
\multirow{7}{*}{\textbf{Fundus2Video}} &
  Knowledge $M$ &
  \ding{55} &
  \ding{55} &
  \multicolumn{2}{c}{\ding{55}} &
  \ding{55} &
  1804.25 &
  0.3225 &
  11.11 &
  0.2213 \\
 &
  Knowledge $M$ &
  \ding{51} &
  \ding{55} &
  \multicolumn{2}{c}{\ding{55}} &
  \ding{55} &
  1611.21 &
  0.3625 &
  11.41 &
  0.2162 \\
 &
  Knowledge $M$ &
  \ding{55} &
  \ding{51} &
  \multicolumn{2}{c}{\ding{55}} &
  \ding{55} &
  1527.94 &
  0.3738 &
  11.76 &
  0.2093 \\
 &
  Knowledge $M$ &
  \ding{55} &
  \ding{55} &
  \multicolumn{2}{c}{\ding{51}} &
  \ding{55} &
  1701.30 &
  0.3694 &
  11.20 &
  0.2133 \\
 &
  Knowledge $M$ &
  \ding{55} &
  \ding{55} &
  \multicolumn{2}{c}{\ding{55}} &
  \ding{51} &
  1664.42 &
  0.3442 &
  11.36 &
  0.2166 \\
 &
  GT Lesion Seg $M$ &
  \ding{51} &
  \ding{51} &
  \multicolumn{2}{c}{\ding{51}} &
  \ding{51} &
  1586.35 &
  0.3688 &
  11.23 &
  0.2136 \\
 &
  Knowledge $M$ &
  \ding{51} &
  \ding{51} &
  \multicolumn{2}{c}{\ding{51}} &
  \ding{51} &
  \textbf{1503.21} &
  \textbf{0.3814} &
  \textbf{11.81} &
  \textbf{0.2001} \\ \hline
\end{tabular}
}

\label{tab:t1}
\end{table*}
\end{center}

\subsubsection{Ablation Studies.}
Additionally, we conduct comprehensive ablation studies to assess the effectiveness of our proposed knowledge mask and related techniques, detailed in the latter part of Table~\ref{tab:t1} and Fig.~\ref{fig:result}. Firstly, we show that our designed mask-enhanced patchNCE loss, knowledge-boost attention, and knowledge-aware discriminators, when combined with mask information, outperform the baseline Fundus2Video. Moreover, our mask-enhanced patchNCE loss yields better results than patchNCE loss alone. Secondly, by replacing the knowledge mask with the ground-truth lesion segmentation mask for comparison, we observe that utilizing our KM-guided techniques can enhance performance even with the lesion segmentation mask. However, our knowledge mask yields better results without the need for additional training or labeling efforts.

\subsubsection{Human Assessment. }
An ophthalmologist reviewed the results of all methods in Table~\ref{tab:t1} and found that our Fundus2Video significantly outperformed others. Then the ophthalmologist conducted a quality assessment of 50 randomly selected FFA videos generated by Fundus2Video from the test set, evaluating them based on their corresponding CF images and ground-truth FFA videos. The evaluation focused on vascular perfusion, lesion dynamics, overall coherence, stability, and presence of artifacts. Scores ranged from 1 to 5, with 1 indicating excellent quality and 5 indicating very poor quality. Our model received a score of 2.12 with a standard deviation of 1.07, indicating good overall quality of the generated videos.

\section{Conclusion}
In this study, we propose Fundus2Video, which pioneers dynamic FFA video generation from static CF images using an autoregressive GAN architecture. With a knowledge mask derived from clinical experience, we enhance focus on dynamic lesion regions, outperforming supervised lesion segmentation masks. Our method incorporates knowledge-boosted attention, knowledge-aware discriminators, and mask-enhanced patchNCE loss to address challenges in lesion generation and pixel misalignment. Fundus2Video emerges as a promising alternative to traditional FFA, surpassing recent state-of-the-art approaches with its non-invasive, intuitive, and dynamic features.
\begin{credits}
\subsubsection{\ackname} The study was supported by the Global STEM Professorship Scheme (P0046113) and the Start-up Fund for RAPs under the Strategic Hiring Scheme (P0048623) from HKSAR. The sponsors or funding organizations had no role in the design or conduct of this research.

\subsubsection{\discintname}
A patent has been filed for this innovation (CN 202410360491.4).
\end{credits}
\bibliographystyle{splncs04} 
\bibliography{ref.bib}
%Author, F.: Article title. Journal \textbf{2}(5), 99--110 (2016)

%Author, F., Author, S.: Title of a proceedings paper. In: Editor,
%F., Editor, S. (eds.) CONFERENCE 2016, LNCS, vol. 9999, pp. 1--13.
%Springer, Heidelberg (2016). \doi{10.10007/1234567890}

%Author, F., Author, S., Author, T.: Book title. 2nd edn. Publisher,

%Author, A.-B.: Contribution title. In: 9th International Proceedings
%on Proceedings, pp. 1--2. Publisher, Location (2010)

%LNCS Homepage, \url{http://www.springer.com/lncs}. Last accessed 4

%\end{thebibliography}
\end{document}